\documentclass[final,5p,times,twocolumn]{elsarticle}

\usepackage{amssymb}
\usepackage{color}
\usepackage{amsmath}
\usepackage{multirow}
\usepackage{booktabs}
\usepackage{bm}

\usepackage[pdfstartview=FitH, colorlinks,
 linkcolor={red!60!black!},
 anchorcolor={green!80!black!},
 urlcolor={blue!80!black!},
 citecolor={blue!50!black!}]{hyperref}
\usepackage[table]{xcolor}

\journal{Physics Letters B}

\begin{document}


\begin{frontmatter}



\title{Chiral three-nucleon force and continuum for dripline nuclei and beyond}


\author[ad1]{Y. Z. Ma}
\author[ad1]{F. R. Xu\corref{cor1}}
\author[ad2]{L. Coraggio}
\author[ad1]{B. S. Hu}
\author[ad1]{J. G. Li}
\author[ad2]{T. Fukui}
\author[ad2]{L. De Angelis}
\author[ad2,ad3]{N. Itaco}
\author[ad2]{A. Gargano}

\address[ad1]{School of Physics, and State Key Laboratory of Nuclear Physics and Technology, Peking University, Beijing 100871, China}
\address[ad2]{Istituto Nazionale di Fisica Nucleare, Complesso Universitario di Monte S. Angelo, Via Cintia, I-80126 Napoli, Italy}
\address[ad3]{Dipartimento di Matematica e Fisica, Universita` degli Studi della Campania “Luigi Vanvitelli”, viale Abramo Lincoln 5 - I-81100 Caserta, Italy}
\cortext[cor1]{frxu@pku.edu.cn}

\begin{abstract}
Three-nucleon force and continuum play important roles in reproducing the properties of atomic nuclei around driplines.
Therefore it is valuable to build up a theoretical framework where both effects can be taken into account to solve the nuclear Schr\"odinger equation.
To this end, in this letter, we have expressed the chiral three-nucleon force within the continuum Berggren representation, so that bound, resonant and continuum states can be treated on an equal footing in the complex-momentum space.
To reduce the model dimension and computational cost, the three-nucleon force is truncated at the normal-ordered two-body level and limited in the $sd$-shell model space, with the residual three-body term being neglected.
We choose neutron-rich oxygen isotopes as the test ground because they have been well studied experimentally, with the neutron dripline determined.
The calculations have been carried out within the Gamow shell model.
The quality of our results in reproducing the properties of oxygen isotopes around the neutron dripline shows the relevance of the interplay between three-nucleon force and the coupling to continuum states.
We also analyze the role played by the chiral three-nucleon force, by dissecting the contributions of the $2\pi$ exchange, $1\pi$ exchange and contact terms.
\end{abstract}


\end{frontmatter}



\section{Introduction}
One of the main challenges in nuclear theory today is to describe weakly-bound and unbound isotopes which locate around the driplines.
These exotic nuclei are also focal points of current and next-generation rare-isotope-beam (RIB) facilities in experimental programs.
Peculiar phenomena, such as halo and narrow resonances, emerge close to the dripline regions, which make the exotic nuclei ideal laboratories to test advanced many-body methods.
The proton-magic oxygen chain is one of the best among these laboratories.
This element is the heaviest one whose neutron dripline position has been pinned down experimentally. 
The interest to study these nuclear systems has been invigorated with
the latest experiments on $^{25-26}$O, which are placed beyond the
dripline.
$^{26}$O ground state has been found barely unbound with two-neutron separation energy of only $-18$ keV \cite{O26_2016}, and an excited state in the unbound resonant isotope $^{25}$O has been observed in a recent experiment \cite{O25_2017}.

Many theoretical works have shown the importance of taking into account the
three-nucleon force (3NF) in nuclear structure calculations with realistic potentials to
reproduce the position of oxygen dripline \cite{Otsuka2010, Hergert2013, Bogner2014, Jansen2014}.
It is shown that 3NF can provide a repulsive
contribution to the binding energies of nuclei, which resolves the overbinding problem in this mass regions.
Besides, it is worth pointing out that the repulsive effect
originates mainly from the long-range $2\pi$ exchange in 3NF \cite{Otsuka2010}.
In weakly-bound and unbound nuclei, it is mandatory to consider also the mixing with continuum states, which is crucial for the understanding of the loose structures of these nuclei.
These nuclear systems have low particle-emission thresholds and hence strong coupling to resonance and continuum.
Resonant and non-resonant continuum states usually have large spatial distributions in their wave functions.
Therefore the bound and spatially-localized harmonic oscillator (HO) basis is no longer effective in expressing the widely-spread resonant and continuum states.
In the last two decades, the conventional shell model (SM) has been extended to include the continuum effect, e.g., shell model embedded in the continuum \cite{Bennaceur1999289, okolowicz2003, Rotureau2005, Volya2005} and continuum-coupled shell model \cite{Tsukiyama2015}.

An elegant treatment of the continuum effect is based on the Berggren method \cite{Berggren1968}, which generalizes one-body Schr{\"o}dinger equation to a complex-momentum plane, creating naturally bound, resonant and continuum single-particle states.
A many-body Hamiltonian may be written in the Berggren basis and diagonalized with complex-energy many-body methods.
Following this approach, it has been developed the so-called Gamow shell model (GSM), which, starting from phenomenological interactions, has proved to be very successful \cite{IdBetan2002, Michel2002, Michel2003, Michel2009}.
The realistic nuclear potential has also been employed to perform GSM
calculations \cite{Tsukiyama2009, Sun2017}, and within the no-core Gamow SM approach \cite{Papadimitriou2013}.
Moreover, also Gamow coupled-cluster calculations have been successfully performed for closed-shell nuclei and nearby \cite{HAGEN2007169}.

\section{Outline of calculations}
In this work, we extend the GSM by introducing a chiral 3NF in the complex momentum Berggren space, where the continuum is included.
More precisely, we start from the chiral two-nucleon force (2NF) derived at next-to-next-to-next-to-leading order (N$^3$LO) by Entem and Machleidt \cite{Machleidt2002} and the three-nucleon force at next-to-next-to-leading order (N$^2$LO) which has been adopted in Refs. \cite{zhuo2018, zhuo2019}.
Then, we derive an effective Hamiltonian for a GSM-with-core calculation \cite{Sun2017} to investigate the structure of oxygen isotopes.

The chiral N$^2$LO 3NF consists of three components, namely the two-pion ($2\pi$) exchange term $V^{(2\pi)}_{\rm 3N}$, the one-pion ($1\pi$) exchange plus contact term $V^{(1\pi)}_{\rm 3N}$ and the contact term $V^{\rm (ct)}_{\rm 3N}$.
It should be pointed out that the low-energy constants (LECs) $c_1$, $c_3$ and $c_4$ \cite{Machleidt2002} which appear in $V^{(2\pi)}_{\rm 3N}$ are the same as in 2NF and have been fixed within the renormalization procedure of the two-body LECs.
However, there are two LECs, $c_D$ and $c_E$, which measure the one-pion exchange and contact terms, respectively.
They cannot be constrained by two-body observables, and need to be determined by reproducing observables in systems with mass number $A>2$.
We adopt the same values of $c_D=-1$ and $c_E =-0.34$ as in Ref. \cite{Navratil2007}.

For the GSM calculation, we choose the doubly-magic $^{16}$O as the inert core, and the $A$-body intrinsic Hamiltonian is broken up into a one-body term $H_0$ and a residual interaction $H_1$ via the introduction of an
auxiliary one-body potential $U$:
\begin{equation}\label{eq:hm1}
 \begin{aligned}
   H&=\sum_{i<j}\frac{(\mathbf{p}_i-\mathbf{p}_j)^2}{2mA}+\hat{V}_{\text{NN}}+\hat{V}_{\text{3N}}\\
    &=\sum_{i=1}^A\left(\frac{p_i^2}{2m}+U\right)
      +\sum^A_{i<j}\left({V}_{\text{NN}}^{(ij)}-U-\frac{p_i^2}{2mA}-\frac{\mathbf{p_i}\cdot \mathbf{p_j}}{mA}\right)\\
    &+\sum^A_{i<j<k}{V}_{\text{3N}}^{(ijk)}\\
    &=H_0+H_1,
 \end{aligned}
\end{equation}
where $H_0=\sum_{i=1}^A(\frac{p_i^2}{2m}+U)$ describes the
independent motion of nucleons.
In the present calculations, we choose the Woods-Saxon (WS) potential
as the auxiliary potential $U$, whose parameters are reported in
Ref. \cite{Sun2017}.
The 3NF matrix element $V^{(ijk)}_{\rm 3N}$ involves six (three initial and three final) states, which increases  the dimension of the shell model drastically.
Particularly for the GSM where continuum states are included, the dimension can be terribly huge due to the inclusions of both 3NF and continuum.
We perform the normal-ordering decomposition \cite{Roth2012} of the 3NF Hamiltonian in order to reduce
the computational complexity.
The final GSM Hamiltonian has a two-body form but includes the 3NF contribution.
It has been shown that the 3NF normal-ordered approximation with neglecting the residual three-body term works well in nuclear structure calculations \cite{Roth2012}.
The normal-ordered two-body term can be written as
\begin{equation}\label{eq:hm2}
 \begin{aligned}
   \hat{V}_{\rm 3N}^{\rm (2B)}
   &=\frac{1}{4}\sum_{\substack{ijkl}}\langle{ij}\vert{V_{\rm 3N}^{\rm (2B)}}\vert{kl}\rangle
      \{{a}^{\dagger}_ia^{\dagger}_ja_la_k\}\\
   &=\frac{1}{4}\sum_{\substack{ijkl}}\sum_{h\in{\rm core}}\langle{ijh}\vert{V_ {\rm 3N}}\vert{klh}\rangle\{{a}^{\dagger}_ia^{\dagger}_ja_la_k\},
  \end{aligned}
\end{equation}
where $\langle{ijh}\vert{V_{\rm 3N}}\vert{klh}\rangle$ and $\langle{ij}\vert{V^{\rm (2B)}_{\rm 3N}}\vert{kl}\rangle$ are the antisymmetrized matrix elements of the 3NF and normal-ordered two-body term, respectively. $a_i^\dagger$ ($a_i$) stands for the particle creation (annihilation) operator with respect to the nontrivial vacuum.
The symbol \{...\} means that the creation and annihilation operators in brackets are normal-ordered.

For the calculations of the {\it sd}-shell nuclei, we choose the doubly magic system $^{16}$O as the core, with its ground-state Slater-determinant as the reference state for the normal-ordering decomposition.
Due to the huge computational cost of generating the 3NF matrix elements, we do not let the states $(i,j,k,l)$ in Eq. (\ref{eq:hm2}) run as freely as in the 2NF calculation, but limit them in the $sd$-shell model space.
This limitation means that 3NF effects from high-lying orbits are missing.
However, it has been shown in Ref. \cite{zhuo2018}, where the same truncation has been adopted, that the effect is not significant since the results obtained within this approximation agree quite reasonably with those of an \textit{ab initio} calculation.
Besides the normal-ordered two-body term, there exist normal-ordered one and zero-body terms \cite{Roth2012}.
In the present work, we adopt the Woods-Saxon Berggren single-particle basis, which is expected to reproduce the experimental single-particle levels in $^{17}$O, thus absorbing the normal ordered one-body term.
The normal-ordered zero-body term, which is just a constant, can be absorbed in the core energy.

We calculate antisymmetrized N$^2$LO 3NF matrix elements in the Jacobi-HO basis in the momentum space and the detail can be found in our previous paper \cite{zhuo2018}.
Then the normal-ordered 3NF two-body matrix elements are added to the N$^3$LO 2NF matrix elements.
The full matrix elements are transformed into the Berggren basis by computing overlaps between the HO and  Berggren basis wave functions (see \cite{Sun2017} for details).
We take a truncation with 22 shells for the HO basis and 20 discretization points for the Berggren continuum contour.
These are the same as in our previous work \cite{Sun2017}, and the convergence has been well tested \cite{Sun2017}.
The complex-energy Berggren representation \cite{Berggren1968} naturally produces bound, resonant and continuum states, which are crucial in the descriptions of weakly-bound and unbound nuclear systems.

Hamiltonian (\ref{eq:hm1}) is intrinsic in the center-of-mass (CoM) frame, but the wave functions are written in the laboratory coordinates.
This means that one should still need to consider the effect from the CoM motion. However, it has been observed that the CoM effect can be neglected for
low-lying states \cite{Sun2017, Hu2019}.

The Berggren basis is generated by the WS potential, including spin-orbit coupling \cite{Dudek1981}.
We adopt the same WS parameters as in Ref. \cite{Sun2017}, obtaining bound $0d_{5/2}$ and $1s_{1/2}$ orbits at energies $-5.31$ MeV and $-3.22$ MeV, respectively, and a resonant $0d_{3/2}$ orbit with energy $\tilde e=1.06-0.09i$ MeV (the eigen energy of a resonant state is written as $\tilde{e}_n=e_n-i\gamma_n/2$, with $\gamma_n$ being the resonance width).
The high-lying orbits $f_{7/2,5/2}$, $ip_{3/2, 1/2}$ ($i{\geq}1$), $g_{9/2,7/2}$, $id_{5/2,3/2}$ ($i{\geq}1$) and $is_{1/2}$ ($i{\geq}2$) are continuum partial waves.
Since $0d_{3/2}$ is a narrow resonant state, playing a crucial role in the description of the $sd$-shell nuclear resonances, the coupling between the $0d_{3/2}$ resonance and $d_{3/2}$ continuum needs to be treated explicitly.
We choose \{$0d_{5/2}$, $1s_{1/2}$, $0d_{3/2}$, $d_{3/2}$-continuum\} as the GSM model space with $^{16}$O as the core.

We construct the effective shell-model interaction in the model space \{$0d_{5/2}$, $1s_{1/2}$, $0d_{3/2}$, $d_{3/2}$-continuum\}, within the framework of the many-body perturbation theory.
More precisely, we employ the so-called Q-box folded-diagram method, which was developed by Kuo and coworkers \cite{Kuo197165}.
The ${\hat Q}$-box folded-diagram method has been extended to the complex-momentum Berggren space, to derive the GSM realistic effective interaction that includes effects from the continuum and core polarization \cite{Sun2017}.
The Berggren model space is nondegenerate.
Therefore we use the extended nondegenerate Kuo–Krenciglowa (EKK) method \cite{Takayanagi201161} for the ${\hat Q}$-box calculation.
The details about the complex-energy ${\hat Q}$-box folded-diagram calculation can be found in Ref. \cite{Sun2017}.
The continuum effect enters into the model through both the effective interaction and active model space.

The complex symmetric non-Hermitian GSM Hamiltonian is diagonalized in the \{$0d_{5/2}$, $1s_{1/2}$, $0d_{3/2}$, $d_{3/2}$-continuum\} space by using the Jacobi-Davidson method in the $m$-scheme.
Due to the presence of the nonresonant continuum, the matrix dimension grows
dramatically when adding valence particles, which is a challenge for diagonalizing in complex space.
In the present calculations, we allow at most three valence particles in the continuum, which can give converged results \cite{Michel2004,Sun2017}.
The WS potential \cite{Sun2017}, based on the universal parameters \cite{Dudek1981}, gives the $0d_{5/2}$ energy lower than the experimental energy by $1.17$ MeV.
As discussed in \cite{Sun2017}, we shift it to reproduce the binding energies of $^{17,18}$O.

\section{Results}
\begin{figure}[!ht]
\begin{center}
\includegraphics[width=0.48\textwidth]{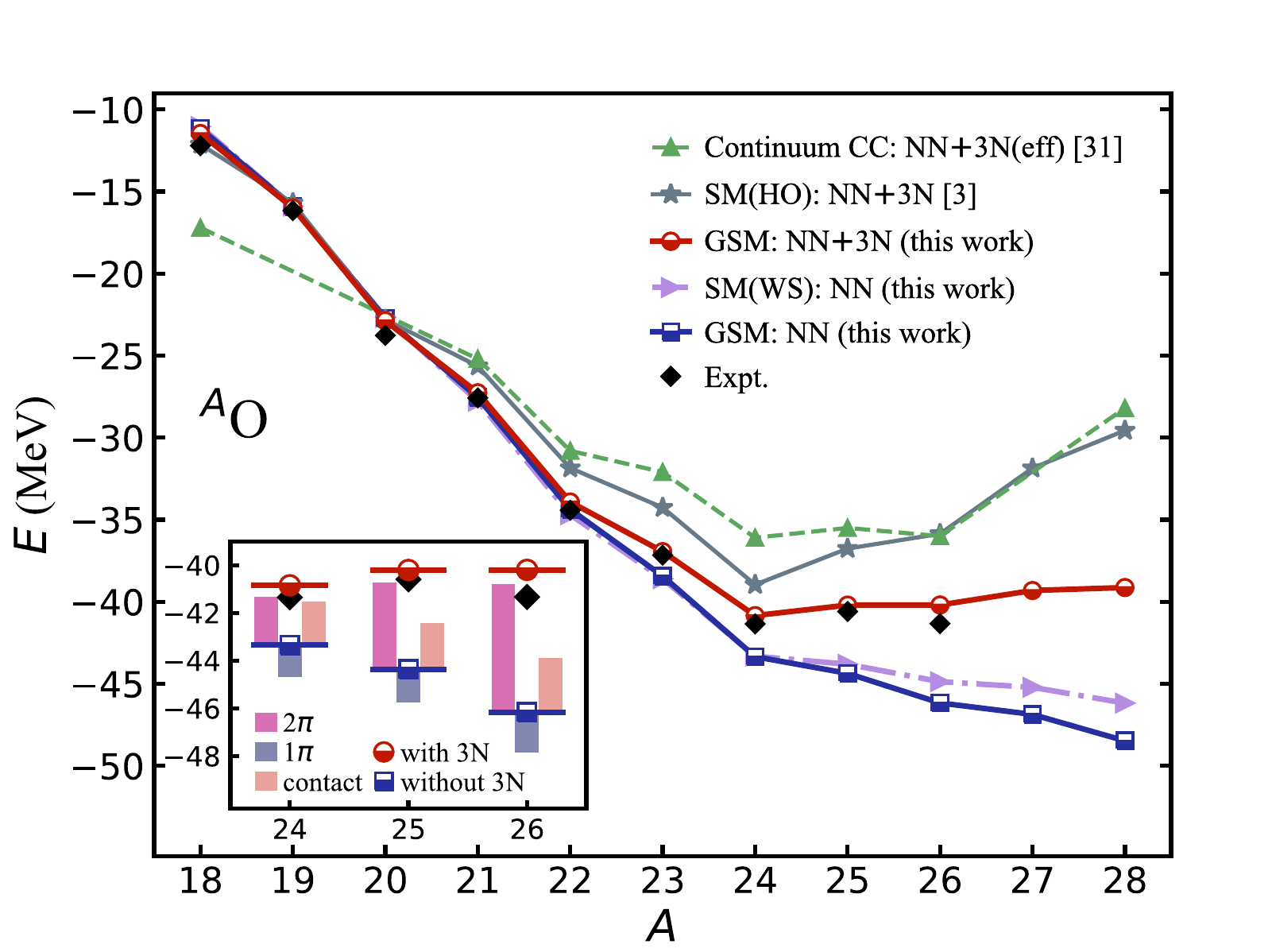}
\caption{\label{fig:BE}Calculated $^{18-28}$O ground-state energies with respect to the $^{16}$O core, compared with experimental data and other calculations: conventional SM(HO) with 3NF but without continuum \cite{Otsuka2010} and continuum coupled cluster (CC) with a density-dependent effective 3NF \cite{Hagen2012}.
SM(HO) and SM(WS) stand for the shell-model calculations performed in the HO and WS bases, respectively.
If not specified, NN and 3N indicate N$^3$LO (NN) and N$^2$LO (NNN), respectively. The inset shows the 3NF contributions from two-pion exchange indicated by ``$2\pi$", one-pion exchange plus contact indicated by ``$1\pi$" and contact term indicated by ``contact".}
\end{center}
\end{figure}

As mentioned in the introduction, we hope that the inclusions of both 3NF and continuum can improve the calculations of weakly-bound and unbound nuclei.
We take neutron-rich oxygen isotopes (up to beyond the neutron dripline) as the test ground.
Figure \ref{fig:BE} shows the calculated $^{18-28}$O ground-state energies with respect to the $^{16}$O core.
The comparison with the results from other calculations shows that both 3NF and continuum play important roles in reproducing the experimental binding energies, particularly in the vicinity of the dripline.
The conventional SM calculations with 3NF but without the continuum \cite{Otsuka2010} reproduce the experimental neutron-dripline nucleus (i.e., $^{24}$O) correctly, but the calculated energies deviate from the experimental data.
The continuum coupled-cluster (CC) calculations with a density-dependent effective 3NF \cite{Hagen2012} provide a similar result, reproducing the dripline qualitatively.
The GSM calculations with only 2NF cannot reproduce the observed neutron dripline, systematically overestimating the ground-state energies.

To see the continuum and 3NF effects within the present model, we have performed the shell-model calculation without the inclusions of the continuum and 3NF.
In this calculation, the WS potential is solved in the HO basis (instead of the Berggren basis), which gives non-continuum discrete WS single-particle states, meaning no continuum included. Within the discrete WS basis and the shell model space $\{0d_{5/2}, 1s_{1/2}, 0d_{3/2}\}$ as in GSM, the shell-model calculation (including $Q$-box folded diagrams) is performed.
The results are shown in Fig. \ref{fig:BE}, indicated by SM(WS) to be distinguished from the SM in the HO basis indicated by SM(HO).
Note that in the SM(HO) calculation of Ref. \cite{Otsuka2010} 3NF is included.
Also, it should be pointed out that in Ref. \cite{Otsuka2010} the two-body N$^3$LO potential has been renormalized by the $V_{\text{low}-k}$ procedure, at variance with the present calculations.
We see that the continuum coupling lowers the ground-state energy while 3NF gives an opposite effect.
The recent experiment observed that $^{26}$O (which is beyond the dripline) is barely unbound with a resonant ground-state energy of only $18$ keV above threshold \cite{O26_2016}.
The present calculation, obtained by taking into account continuum and 3NF effects, reproduces the experimental observation.
Table \ref{tab:table1} lists the calculated two-neutron separation energies for the dripline nucleus $^{24}$O and beyond.
We see that the GSM with the inclusion of 3NF improves the calculations of the separation energies significantly, compared with data.
\begin{table}[h]\small
\center
\setlength{\belowcaptionskip}{5pt}%
\caption{\label{tab:table1}Two-neutron separation energies $S_{\rm 2n}$, calculated by GSM with only two-body (NN) force at N$^3$LO and with the inclusion of three-body (3N) force at N$^2$LO, compared with data \cite{Audi2012, O26_2016}.}
\renewcommand\arraystretch{1.2}
\setlength{\tabcolsep}{4.2mm}
\begin{tabular}{llll}
\hline
\hline
$S_{2n}$(MeV) & NN  & NN+3N  & Expt. \\ \hline
$^{24}$O    & $9.110$   & $6.924$ & $6.92$ \\
$^{25}$O    & $6.254$   & $3.259$ & $3.453$ \\
$^{26}$O    & $3.362$   & $-0.648$ & $-0.018$ \\
\hline
\hline
\end{tabular}
\end{table}

\begin{figure}[!ht]
\begin{center}
\includegraphics[width=0.46\textwidth]{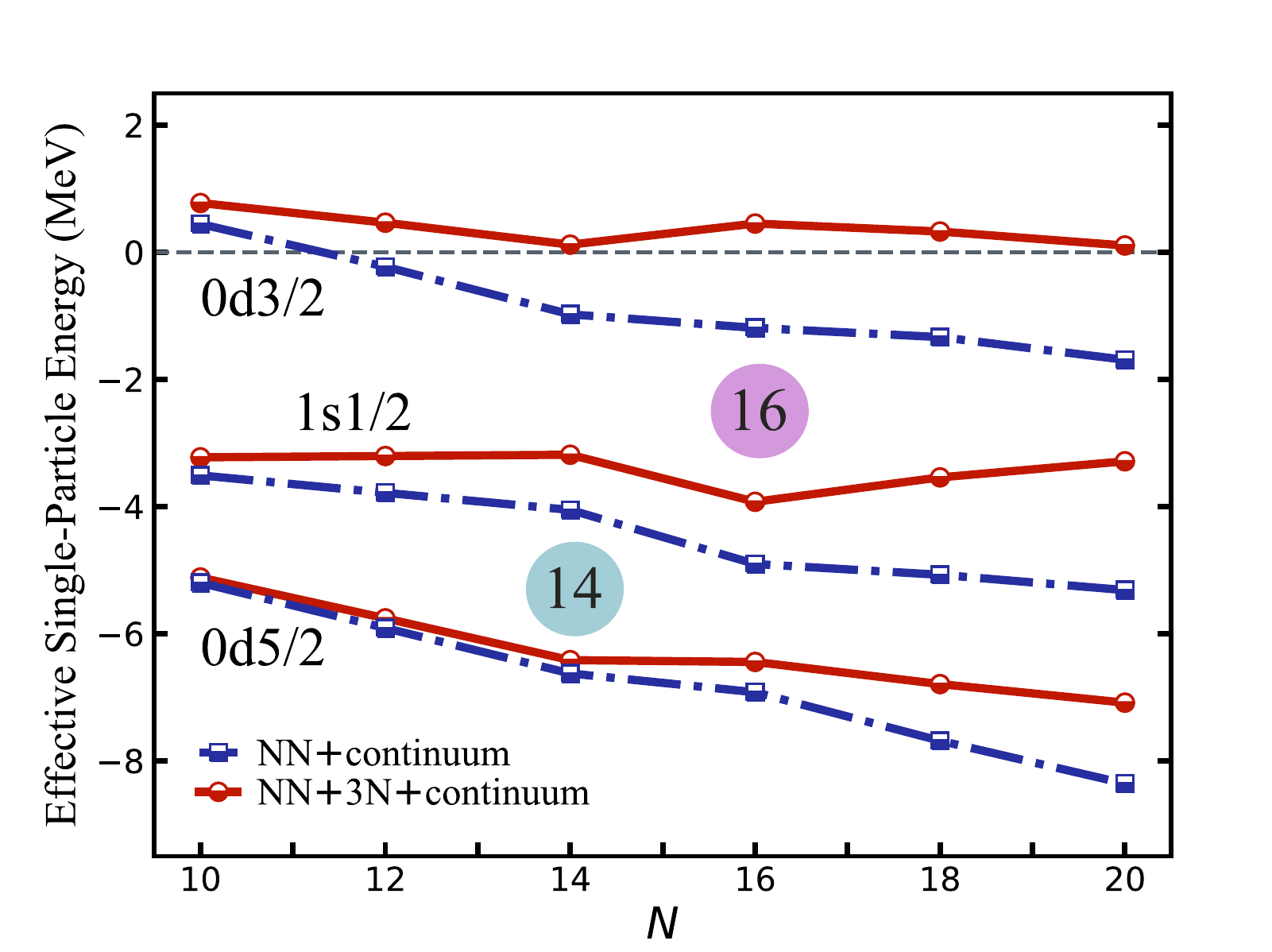}
\caption{\label{fig:ESPE} Neutron effective single-particle energies in the oxygen chain. The red solid and blue dot-dash curves are for the calculations with and without 3NF, respectively. The numbers of $N=14, 16$ indicate the neutron sub-shell closures.}
\end{center}
\end{figure}

As shown in Fig. \ref{fig:BE}, the results of GSM calculations provide a clear difference in the reproduction of ground-state energies when including or excluding 3NF.
This difference enlarges with the increase of the number of valence neutrons.
The 3NF gives strong repulsive contributions to the ground-state energies of the nuclei in the vicinity of the dripline.
Figure \ref{fig:ESPE} shows the neutron effective single-particle energies (ESPE), see \cite{zhuo2019} for the definition of the ESPE.
We see that the neutron ESPE's calculated without 3NF (blue curves in Fig. \ref{fig:ESPE}) drop persistently with increasing the neutron number, while the 3NF pushes up the single-particle orbitals (red curves).
In particular, the $0d_{3/2}$ and $1s_{1/2}$ ESPEs are pushed up significantly by the 3NF, while the 3NF does not change so much the $0d_{5/2}$ ESPE for $N\leq 14$.
For the oxygen isotopes with $N\leq 14$ (i.e., lighter than $^{23}$O), the $1s_{1/2}$ and $0d_{3/2}$ are almost empty. Hence their binding energies are less affected by the 3NF, see Fig. \ref{fig:BE}.
From $^{23}$O ($N=15$), the $1s_{1/2}$ and $0d_{3/2}$ orbitals start to be occupied.
Therefore, the 3NF effects become significant, which leads to remarkable improvements in the reproduction of the observed binding energies and two-neutron separation energies for $^{23-26}$O (see Fig. \ref{fig:BE} and Table \ref{tab:table1}).
Moreover, in Fig. \ref{fig:ESPE}, we see that the 3NF enhances the neutron sub-shell closures at $N=14$ and $16$.

In the inset of Fig. \ref{fig:BE}, we dissect the 3NF effect in $^{24-26}$O, where the 3NF effects are significant.
The two-pion exchange $V^{(2\pi)}_{\rm 3N}$ and the contact term $V^{(ct)}_{\rm 3N}$ supply repulsive contributions to the binding energy, indicated by bars above the 2NF GSM energy, while the one-pion exchange plus contact term $V^{(1\pi)}_{\rm 3N}$ produces an attractive contribution, displayed by a bar below the 2NF GSM energy.
We also see that, besides the two-pion exchange $V^{(2\pi)}_{\rm 3N}$ which corresponds to the long-range 3NF interaction, the one-pion exchange and contact terms also have non-negligible contributions.
The $V^{(1\pi)}_{\rm 3N}$ and $V^{(ct)}_{{\rm 3N}}$ contributions are very close in absolute value but opposite in signs.
Consequently, their net effect is almost canceled out, implying that the two-pion exchange is responsible for the 3NF strong repulsive effect.
As can be seen in the inset of Fig. \ref{fig:BE}, the role of the two-pion exchange becomes more important when increasing the neutron number, while the contributions from the one-pion exchange and contact components are almost unchanged.

\begin{figure}[!ht]
\begin{center}
\includegraphics[width=0.46\textwidth]{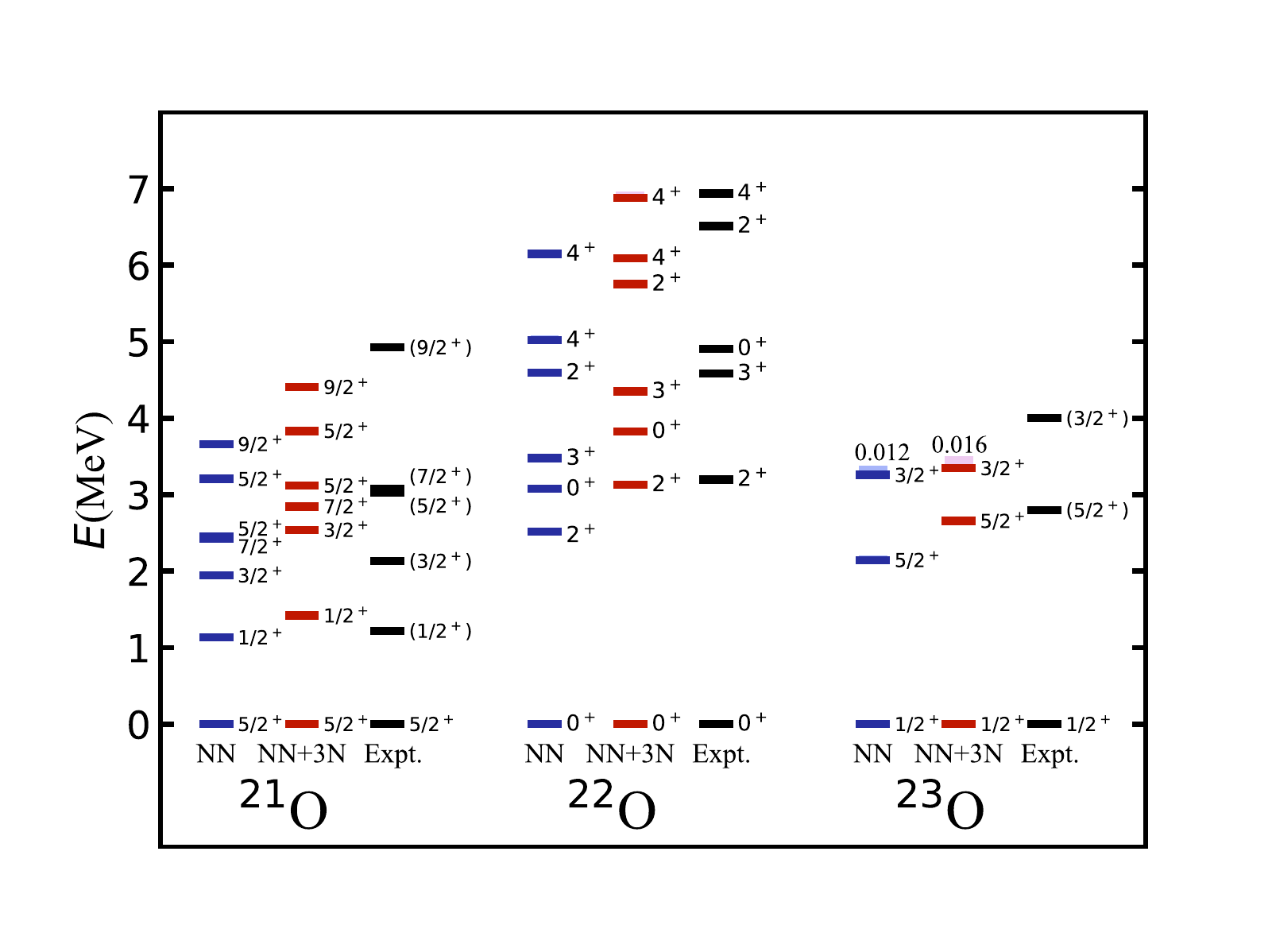}
\caption{\label{fig:O212223}Calculated spectra for $^{21-23}$O by the GSM with 2NF (NN) only and with 3NF included (NN+3N), compared with available data \cite{BASUNIA201569,Elekes2007,FIRESTONE20151,Schiller2007}. The resonant state is indicated by shading, with the resonance width (in MeV) given by the number above the level.}
\end{center}
\end{figure}

The GSM can describe both bound and unbound states on equal footing. Figure \ref{fig:O212223} displays the calculated spectra of the neutron-rich bound isotopes $^{21-23}$O.
We see that 3NF improves agreements with experimental data.
The $N=14$ sub-shell closure at $^{22}$O is clearly seen with a large excitation energy of the $2_1^+$ state.

\begin{figure}[!ht]
\begin{center}
\includegraphics[width=0.46\textwidth]{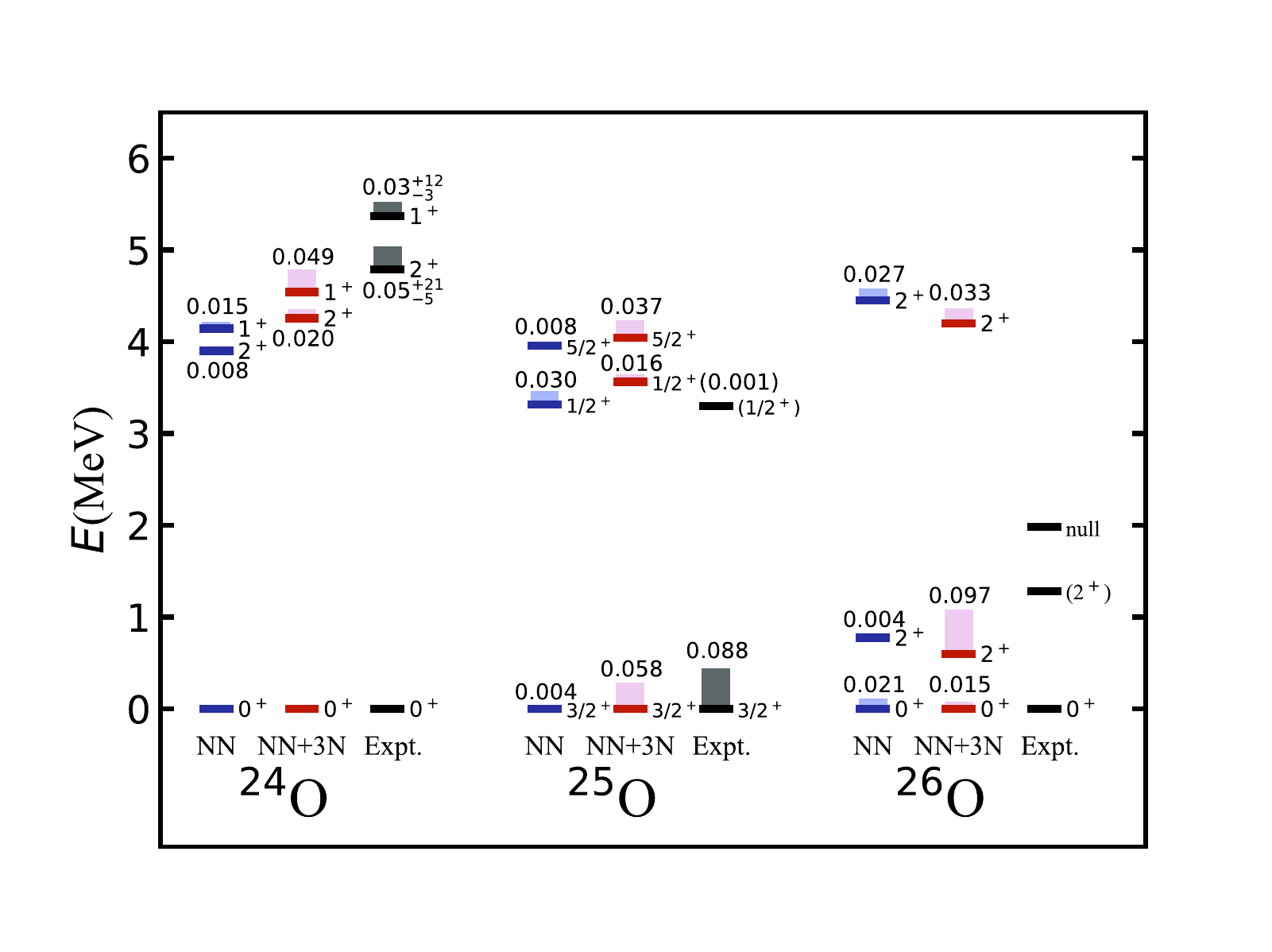}
\caption{\label{fig:O242526}Similar to Fig. \ref{fig:O212223}, but for $^{24-26}$O. The experimental data are taken from \cite{Hoffman200917, O25_2017, O26_2012, O26_2016}.}
\end{center}
\end{figure}

Our interests are in weakly-bound and unbound nuclei.
Figure \ref{fig:O242526} shows the spectroscopic calculations with and without the 3NF effects, compared with existing experimental observations, for the oxygen isotopes near the neutron dripline.
In $^{24}$O, the observed resonant excited $2^+_1$ and $1^+_1$ states are reproduced, and the 3NF improves the calculations in both the excitation energies and resonance widths.
In $^{25}$O, the GSM calculation with the N$^3$LO 2NF gives over-bound binding energy, while the inclusion of the N$^2$LO 3NF describes well the unbound resonant property of the ground state with the resonance width agreeing well with the  experimental measurement.
A new excited state with a possible configuration of $J^{\pi}=1/2^+$ was reported recently in the experiment \cite{O25_2017}.
The present calculations support the experimental suggestion.

As discussed above, the GSM calculation with 3NF can well describe the ground state of the observed unbound $^{26}$O beyond the dripline.
As regards the unbound $^{26}$O, in a recent experimental work, a ($2^+$) state has been observed at an excitation energy of $1.28$ MeV, while the experimental resolution has not been able to establish the resonance widths.
Our results with 3NF provide, besides an unbound ground state with a resonance width of $15$ keV, a $2^+$ state that is lower in energy than the observed one and whose resonance width is $97$ keV.
We also predict a second $2^+$ state at $\sim 4$ MeV with a width of $33$ keV.
It should be mentioned that $^{26}$O has been calculated in the
framework of GSM by using a phenomenological two-body residual interaction \cite{Fossez2017}.
The results in Ref. \cite{Fossez2017} predicts a barely bound ground state and a $2^+_1$ state at $E_x \approx$ $1.08$ MeV with a resonance width of $\sim$ $27$ keV.
Therefore, it is worth pointing out that the above results evidence that the interplay between continuum and 3NF effects reveals itself crucial to reproduce the oxygen dripline correctly.

\section{Conclusions}
In conclusion, we have been successful in extending the chiral N$^2$LO three-body interaction to the complex-momentum Berggren space in which the resonance and continuum are included.
To reduce the computation task, the 3NF is normal-ordered.
With the chiral N$^3$LO 2NF and N$^2$LO 3NF, we have performed the Gamow shell-model calculations for neutron-rich oxygen isotopes as the test ground.
The calculations with the inclusions of both 3NF and continuum reproduce the dripline position and the unbound properties of the nuclei beyond the dripline.
The present calculations explain well the experimental resonance widths of the $^{24}$O excited states and predict the particle-emission widths for other resonant states of the isotopes.
The 3NF two-pion exchange and the contact term have repulsive contributions to binding energies, while the one-pion exchange has an attractive contribution.
The one-pion exchange and the contact term have similar strengths but opposite effects (attractive and repulsive, respectively).
The contribution of the two-pion exchange increases with increasing the neutron number in the oxygen chain, playing a crucial role in the descriptions of the data.
The 3NF significantly pushes up the $0d_{3/2}$ and $1s_{1/2}$ orbits, which are heavily occupied in the isotopes beyond the dripline $^{24}$O.
Therefore, 3NF becomes important in the dripline region of the ${\it sd}$ shell.
The recent experimental observation of the barely unbound $^{26}$O is reproduced reasonably, with a small unbound two-neutron separation energy, which is close to the experimental datum.
The comparison with the other calculations has evidenced the importance of including both 3NF and continuum effects for the class of isotopes under investigation.

\section{Acknowledgements}
Valuable discussions with Nicolas Michel, Simin Wang and Zhonghao Sun are gratefully acknowledged.
This work has been supported by the National Key R\&D Program of China under Grant No.2018YFA0404401, the National Natural Science Foundation of China under Grants No.11835001, No. 11921006, No. 11575007, and No.11847203; and the CUSTIPEN (China-US Theory Institute for Physics with Exotic Nuclei) funded by the US Department of Energy, Office of science under Grant No. DE-SC0009971.
We acknowledge the High-performance Computing Platform of Peking University for providing computational resources.



  \bibliographystyle{elsarticle-num_noURL}
  \bibliography{elsarticle-zhuo}





\end{document}